\documentclass[11pt]{article} 

\usepackage[margin=1in]{geometry}

\usepackage{graphicx} 
\usepackage{dcolumn} 
\usepackage{bm,amsmath,amsthm,amssymb,amsfonts}
\usepackage{physics}
\usepackage{palatino}
\usepackage{mathtools}
\usepackage{algorithm}
\usepackage{algpseudocode}
\usepackage{dsfont}
\usepackage{xcolor}
\usepackage{comment} 
\usepackage{rotating}
\usepackage{tikz}
\usepackage{siunitx}
\usepackage[outline]{contour} 
\usepackage{hyperref}       
\usepackage{url}            
\usepackage{nicefrac}       
\usepackage{microtype}      
\usepackage{cleveref}       
\usepackage{graphicx}
\usepackage{natbib}
\usepackage{doi} 
\usepackage{verbatim}
\usepackage{subfig}
\usepackage{cancel}

\usetikzlibrary{quantikz}
\usetikzlibrary{angles,quotes} 
\contourlength{1.3pt}
\tikzset{>=latex} 
\tikzstyle{vector}=[->,very thick]
\tikzstyle{mydashed}=[dash pattern=on 2pt off 2pt]
 
\tikzstyle{startstop} = [rectangle, rounded corners, minimum width=2.5cm, minimum height=1cm, text centered, draw=black, fill=red!30, text width=2cm]
\tikzstyle{process} = [rectangle, minimum width=2.5cm, minimum height=1cm, text centered, draw=black, fill=blue!30, text width=2cm]
\tikzstyle{arrow} = [thick,->,>=stealth]

\theoremstyle{definition}
\newtheorem{definition}{Definition}[section]

\newcommand{\A}{\mathcal{A}}
\newcommand{\B}{\mathcal{B}}

\newcommand{\K}{\mathcal{K}}

\newcommand{\X}{\mathcal{X}}
 
\newcommand{\Ocomp}{\mathcal{O}}

\newcommand{\Pos}{\mathrm{Pos}}

\newcommand{\Herm}{\mathrm{Herm}}
\newcommand{\Sep}{\mathrm{Sep}}
\newcommand{\SepD}{\mathrm{SepD}}

\newcommand{\kb}[1]{| #1 \rangle \langle #1 |}

\newcommand{\id}{\mathds{1}}

\newcommand{\Cspace}{\mathbb{C}}

\newcommand{\Sset}{\mathcal{S}}

\newcommand{\hsip}[2]{\left\langle #1, #2 \right\rangle}
\renewcommand{\eqref}[1]{Eq.~(\ref{#1})}  

\usepackage{color}

\definecolor{KB}{rgb}{0.4,0.3,0.9}
 
\definecolor{THc}{rgb}{0.9,0.3,0.2}

\begin{document}

\title{{Quantum heuristics for linear optimization over large separable operators}}

\author{
Ankith Mohan
\thanks{
Virginia Tech, 
Blacksburg, VA, USA $24061$. 
\texttt{\{ankithmo, sikora\}@vt.edu} 
}
\and
Tobias Haug
\thanks{
Quantum Research Center, Technology Innovation Institute, Abu Dhabi, UAE.
\texttt{tobias.haug@u.nus.edu}
}
\and
Kishor Bharti
\thanks{
Agency for Science, Technology and Research, Singapore $138632$. 
\texttt{kishor.bharti1@gmail.com}
}
\and
Jamie Sikora
\footnotemark[1]
} 

\date{\quad \\\today} 

\maketitle

\begin{abstract}  
Optimizing over separable quantum objects is challenging for two key reasons: determining separability is NP-hard, and the dimensionality of the problem grows exponentially with the number of qubits.
We address both challenges by introducing a heuristic algorithm that leverages a quantum co-processor to significantly reduce the problem’s dimensionality. 
We then numerically demonstrate that see-saw-type optimization performs well in lower-dimensional settings.
A notable feature of our approach is that it yields feasible solutions, not just bounds on the optimal value, in contrast to many outer-approximation-based methods.
We apply our method to the problem of finding separable states with minimal energy for a given Hamiltonian and use this to define an entanglement measure for its ground space. 
Finally, we demonstrate how our approach can approximate the separable ground energy of Hamiltonians up to 28 qubits.
\end{abstract} 

\bigskip 

\section{Introduction}\label{sec:intro}  
Entanglement is at the heart of many quantum problems and applications~\cite{horodecki2009quantum}.   
Even though it has been studied for a long time now, entanglement is still a mysterious resource and understanding its power and limitations is an interesting area of research.  
Given a quantum state, determining whether it is entangled or separable (i.e., not entangled) is called the \emph{separability problem}.  
More precisely, for a density operator acting on the bipartite space $\A \otimes \B$, we wish to decide whether it belongs to the set of separable states, defined as 
\begin{equation} 
\SepD(\A: \B) = \mathrm{conv} \; \{ \rho \otimes \sigma \} 
\label{eq:SepIntro}
\end{equation} 
where $\mathrm{conv}$ denotes the convex hull (the set of all convex combinations), $\rho$ ranges over density operators on $\A$, and $\sigma$ ranges over density operators on $\B$. 
This problem is known to be NP-hard in general~\cite{gurvits2004classical,gharibian2010strong}. 
If $\dim(\A) \cdot \dim(\B) \leq 6$, then it turns out that this problem has a finite-sized semidefinite programming formulation~\cite{horodecki1996necessary} while for all other cases, it was shown that no finite-sized semidefinite program exists~\cite{fawzi2021set}. 

In this work, we also look at the set of separable \emph{operators} which is  defined as  
\begin{equation} 
\Sep(\A: \B) = \mathrm{conv} \; \{ X \otimes Y \} 
\label{eq:Sep}
\end{equation} 
where $X$ is a positive semidefinite operator acting on $\A$ and $Y$ is a positive semidefinite operator acting on $\B$. 
We note that $\Sep(\A: \B)$ and $\SepD(\A: \B)$ differ only in the fact that the latter has a unit trace constraint. 

A general linear optimization over the set of separable operators takes the form 
\begin{equation} \label{eq:Sep_opt_operator} 
\text{maximize } \Big\{ \hsip{X}{C} : \Xi(X) = B, X \in \Sep(\A:\B) \Big\} 
\end{equation} 
where $C$ and $B$ are Hermitian matrices, $\Xi$ is a hermiticity-preserving linear map, and $\hsip{A}{B}$ denotes the Hilbert-Schmidt inner product between two operators $A$ and $B$, which is defined as $\Tr(A^* B)$.  
While we consider the general problem above in this work, we often focus on the special case of linear optimization over the set of separable \emph{states} taking the form 
\begin{equation} \label{eq:Sep_opt} 
\text{maximize } \Big\{ \hsip{\tau}{C} : \tau \in \SepD(\A:\B) \Big\}.  
\end{equation} 
Despite having a diverse array of applications~\cite{ioannou2007computational, brandao2011quasipolynomial, gharibian2013qma, chailloux2012complexity, hayden2013two, li2014quantum, consiglio2022variational, jeronimo2023power}, 
this problem is hard for two key reasons.   
For one, even if one were to exhibit a purported optimal solution, verifying its separability could be (NP) hard. 
Also, it suffers from the curse of dimensionality. 
In particular, when $\A$ and $\B$ represent $n$-qubit spaces, then $\tau \in \SepD$ has dimension $2^{2n}$ which renders many numerical computations infeasible. 

In this work, we address both of these issues by combining semidefinite programming techniques with quantum heuristics. 
Firstly, by using semidefinite programming-based algorithms, one can solve small instances (i.e., when $\dim(\A)$ and $\dim(\B)$ are small) with reasonable levels of success. 
Secondly, we introduce a heuristic algorithm to  reduce the dimension of the optimization problem. 
We discuss when this pre-processing step can be run on a quantum co-processor, thereby alleviating the curse of dimensionality. 
In a sense, we do not need to do any vector-represented calculations, the interactions are done on a quantum mechanical level and the desired quantities can be inferred from measurement statistics. 
We combine both of these methods, allowing us to optimize over separable states of large dimensions. 

We note that even though our algorithm is heuristics-based, it maintains an important feature which is that it always returns a feasible solution.\footnote{Interestingly, even if the quantum processing part of our method is noisy, it still returns a feasible solution.} 
Thus, in the event that our algorithm does not perform well,\footnote{Due to assumed complexity theoretical barriers (P $\neq$ NP), this is likely the case eventually for any such algorithm.} it still exhibits a feasible solution which could still be of value depending on the application. 

As an application, we approximate the separable state that has the greatest energy with respect to a fixed Hamiltonian, whose energy we call the \emph{separable ground energy}.\footnote{Note that since we are maximizing throughout the discussions in this work, we think of the largest eigenvalue as the ground energy. 
This convention is without loss of generality since one can replace the Hamiltonian $H$ with $-H$ if one prefers to minimize.} 
We show that this approximation can be used to characterize the amount of entanglement in the groundspace of the Hamiltonian. 
We give a full study of our algorithm for the case of the one-dimensional Ising Hamiltonian with numerics for up to $28$ qubits. 

\subsection{Related work for optimizing over separable states}  

An algorithm was proposed in \cite{brandao2011quasipolynomial} that solves the optimization problem~(\ref{eq:Sep_opt}) in time 
\begin{equation} 
\exp(\Ocomp(\delta^{-2}\ \log(\dim \A)\ \log(\dim \B)\ \| C \|_F^2)), 
\end{equation} 
for additive error $\delta$ which is based on \emph{symmetric extensions} introduced in~\cite{doherty2002SymExt}. 
Another approach that combines KKT conditions with symmetric extensions was studied in~\cite{harrow2017improved}. 
In \cite{shi2015epsilon}, the authors propose a quasi-polynomial time algorithm which takes as input the Hermitian matrix $C$ and its decomposition $C = \sum_{i=1}^M C_i^1 \otimes C_i^2$ for small $M$.
Then optimizing over product states $\hsip{\rho \otimes \sigma}{C}$ becomes $\sum_{i=1}^M \hsip{\rho}{C_i^1} \hsip{\sigma}{C_i^2}$. 
By modifying the construction of $\varepsilon$-nets (\cite{shi2015epsilon}, Definition $1$) using the multiplicative matrix weights (MMW) update method~\cite{arora2012multiplicative}, we can enumerate all possible values of $\hsip{C_i^1}{\rho}$ for all values of $i$. 
Thus the problem can be solved efficiently when $M$ is small. 
Moreover, when $C$ is positive semidefinite, they propose another algorithm which combines this $\varepsilon$-net approach with Schmidt decompositions to approximate the optimal value of problem~(\ref{eq:Sep_opt})  
in time 
\begin{equation} 
\exp(\Ocomp(\log(d) + \delta^{-2} \|C\|_F^2 \ln(\|C\|_F/\delta)))  
\end{equation} 
where $d = \dim(\A) = \dim(\B)$.  
There are also several variational approaches to solving this problem~\cite{consiglio2022variational, philip2024schrodinger}. 

\paragraph{Comparisons to our algorithm.} 
One issue that still arises in working with the hierarchy of symmetric extensions, or other popular outer approximations to $\Sep$ (such as PPT (Positive-Partial Transpose) operators), is the curse of dimensionality. 
The sizes of the variables in the semidefinite programs still scale exponentially, and this problem is only exacerbated at higher levels. 
While these algorithms are great in general, their size is still an issue, especially with current numerical solvers.  
We do note that these algorithms, as well as others, can be combined with our dimension-reducing quantum heuristics though. 

Our approach has some similarities to \cite{shi2015epsilon}. 
We look at general $C$ matrices as well as those that can be decomposed as a sum of tensor products. 
While their approach uses MMW, we use dimension-reduction techniques to get the size of the problem smaller, then generalized eigensolvers, see Section~\ref{sec:small_SEP}. 
Thus, the approaches are complimentary in this setting. 
We do note that since our approach is heuristics-based, it can be used in general settings, sometimes even when the size of $C$ is exponentially large, as discussed in Section~\ref{sec:big_sep}. 

Comparing to the previously-mentioned variational algorithms, our algorithm promises to give feasible solutions whereas this is not always the case in  variational algorithms. 
This could be an important feature, depending on the application.
Unlike variational algorithms, our approach does not involve the optimization of parameters in any parametric quantum circuit, thereby circumventing the barren plateau problem~\cite{mcclean2018barren} by design. Furthermore, the absence of a classical-quantum feedback loop, characteristic of Variational Quantum Algorithms, significantly reduces the overhead associated with quantum measurements.

\subsection{Related work for optimizing over separable ground states of Hamiltonians} 
The mean-field method is a powerful class of approximations for complex interacting Hamiltonians, reducing them to effective one-body problems that are easier to analyze. A prominent example is the Hartree-Fock approach for fermionic Hamiltonians, which simplifies the interaction terms to one-body fermionic operators~\cite{google2020hartree}. The core idea is to replace the original Hamiltonian 
$H$ with a simpler, mean-field Hamiltonian $H_{\text{MF}}$. By construction, the ground state of $H_{\text{MF}}$ is a separable state (i.e., a product state of single-particle wave functions). 
A folklore result is that the computation of the ground-state energy of a mean-field Hamiltonian $H$ is equivalent to solving a problem of the form (\ref{eq:Sep_opt}). 
However, the mean-field ansatz is highly constrained. This assumed separability (often defined in a specific, potentially non-local basis corresponding to the particle wave functions) means that the ground state of the mean-field Hamiltonian frequently lacks important correlations present in the true ground state, potentially rendering it an inadequate approximation.

\paragraph{Comparisons to our algorithm.} Our algorithm is distinct from the mean-field approach in a crucial way. The mean-field method first approximates the Hamiltonian to get $H_{\text{MF}}$ and then finds its exact (separable) ground state. In contrast, our approach directly finds an approximate separable ground state for the original Hamiltonian $H$. The separable states obtained by these two different methods are generally not the same. Our method's flexibility stems from an ansatz that can systematically incorporate higher-order terms, which are explicitly neglected in standard mean-field and Hartree-Fock approximations. In fact, our approach can be built directly on top of these methods. For instance, the Hartree-Fock ground state can serve as the initial reference state for our heuristic, which can then be improved by including higher-order Pauli terms that go beyond the mean-field limitations. 
  
\subsection{Paper organization} 

In the next section we study optimization-based algorithms for solving (small) separability problems and examine numerically how well they perform. 
We introduce our quantum heuristics in Section~\ref{sec:big_sep} and, as a concrete application, apply it to finding separable ground states of Hamiltonians in Section~\ref{sec:Hammy_intro}.  
In particular, in that application we study the effect the separability constraint has on the optimization problem (or in the context of Hamiltonians, its ground energy). 
We also examine how well the algorithm performs as we modify the ansatz in our heuristic. 
Lastly, we study the computational limits of our methods in general. 
     

\section{Semidefinite programming-based algorithms for the separability problem, a quick discussion}\label{sec:small_SEP} 

We say that $\K$ is a convex cone if it is convex and also closed under nonnegative scaling, meaning that if $x \in \K$ then $\lambda x \in \K$ for all $\lambda \geq 0$. 
A cone program is the study of optimizing a linear function over a variable in a convex cone $\K$ subject to affine constraints. 
In standard form, we can write one as 
\begin{equation} 
\text{maximize } \Big\{ \hsip{X}{A} : \Xi(X) = B, \; X \in \K \Big\} 
\end{equation} 
where $\Xi$ is a hermiticity-preserving linear map, and $A$ and $B$ are Hermitian. 
In this work, we are concerned primarily with the cone of \emph{separable operators}, i.e., when $\K = \Sep$, but also when $\K$ is the set of positive semidefinite operators, denoted by $\Pos(\X)$ when acting on a complex Euclidean space $\X$.  
The above optimization in this case is called \emph{semidefinite programming} and is of great interest in many areas including combinatorial optimization~\cite{tunccel2016polyhedral} and quantum information~\cite{watrous2018theory} and we refer to a specific instance as a semidefinite program (SDP). 
While optimizing over separable operators is indeed intractable, one can often solve SDPs efficiently. 
Current numerical solvers for solving SDPs exist such as CVX~\cite{grant2014cvx}, SeDuMi~\cite{sturm1999using}, SDPT3~\cite{toh1999sdpt3} and Mosek~\cite{andersen2000mosek}. 

Since SDPs can often be solved efficiently, they are a popular choice for approximating convex sets which are difficult to deal with, a great example being $\Sep$. 


Let us consider again the optimization problem~(\ref{eq:Sep_opt}).  
Notice that if we have the optimal solution $\tau$, which always exists by compactness of the set $\SepD$, then a simple convexity argument\footnote{If an optimal solution is a convex combination of feasible solutions, then one of those feasible solutions has objective function value at least as good as the original optimal solution.} shows that there exists pure states $\ket{\psi}$ and $\ket{\phi}$ such that $\kb{\psi} \otimes \kb{\phi}$ is an optimal solution.   
Therefore, we can optimize over product states instead of all of $\SepD$. 
Under this restriction, we can rewrite the problem~(\ref{eq:Sep_opt}) as 
\begin{equation} \label{eq:JohnWick4}
\text{maximize } \Big\{ \hsip{\rho \otimes \sigma}{C} : \rho \in D(\A), \; \sigma \in D(\B) \Big\},   
\end{equation}  
where we have denoted the set of \emph{density matrices} acting on $\X$ as $D(\X)$. 
(Technically, we can assume $\rho$ and $\sigma$ to be pure if it helps, but we do not need to assume this.)  
Since the optimization problem~(\ref{eq:JohnWick4}) is quadratic, it still remains hard in general. 
However, there are folklore algorithms called \emph{see-saw} which often perform well in practice (and are a special case of coordinate descent algorithms~\cite{wright2015coordinate}).
(Note that no algorithm will always perform well due to NP complexity barriers.)   
The see-saw pseudo-algorithm is stated below for this setting, noting the notation $\Herm(\X)$ as the set of Hermitian operators acting on $\X$.  

\begin{algorithm}[htpb]
    \caption{See-saw}\label{algo:see-saw}
    \begin{algorithmic}[1]
        \State\textbf{Input:} $\hat{\rho} \in D(\A), C \in \Herm(\A \otimes \B)$ 
        \Repeat
            \State Solve for optimal $\hat{\sigma} \in D(\B)$: $\mathrm{maximize} \hsip{\hat{\rho} \otimes \sigma}{C}$. 
            \State Solve for optimal $\hat{\rho} \in D(\A)$: $\mathrm{maximize} \hsip{\rho \otimes \hat{\sigma}}{C}$. 
        \Until{Progress is not being made.}
        \State\Return $\hat{\rho}$ and $\hat{\sigma}$. 
    \end{algorithmic} 
\end{algorithm} 
  
This approach generally works well, but it could depend on how the initial state $\hat{\rho}$ is chosen. 
Of course, if one starts with $\hat{\rho}$ from an optimal solution $\rho \otimes \sigma$, then this algorithm terminates immediately with an optimal solution. 
However, due to the curvature of the function, it can easily get stuck in a local maximum, rendering it difficult to find a global maximum without a restart. 
We consider three choices of initial states in this work, the maximally mixed state $\hat{\rho} = \frac{1}{\dim(\A)}\id_\A$, the uniform superposition state $\hat{\rho} = \kb{\psi}$ where $\ket{\psi} = \frac{1}{\sqrt{\dim(\A)}} \sum_{i} \ket{i}$, and $100$ randomly chosen states.  

An important aspect of the see-saw algorithm is that it gives a feasible solution, in the sense that $\hat{\rho} \otimes \hat{\sigma}$ is always separable. 
Thus, this gives us a \emph{lower bound} on the optimal value and a corresponding feasible solution. 
We note that many SDP-based algorithms for approximating the separable set are \emph{outer approximations} which do not always have this feature of yielding a feasible solution (such as the PPT criterion~\cite{horodecki1996necessary, peres1996separability}, symmetric extensions~\cite{doherty2002SymExt, chen2014symmetric, li2019symmetric}, and the realignment criterion~\cite{chen2003matrix}). 
Therefore, this is a nice feature of the see-saw algorithm. 


\subsection{A bottleneck in the see-saw algorithm} \label{sec:faster_see-saw} 

Recall that in see-saw we wish to alternately optimize over $\rho$ and $\sigma$. 
Concentrating on optimizing over $\sigma$ for now, we can rewrite $\hsip{\hat{\rho} \otimes \sigma}{C}$ as $\hsip{\sigma}{C_\B}$, where $C_\B = \Tr_\A[(\hat{\rho} \otimes \id)C] \in \Herm(\B)$ and $\Tr_{\A}$ is the partial trace over $\A$. 
Thus, we are effectively solving the optimization problem 
\begin{equation}
    \text{maximize } \Big\{ \hsip{\sigma}{C_\B} : \sigma \in D(\B) \Big\}
\end{equation}
whose solution corresponds to the principal eigenvector of $C_\B$. 
Similarly, $\hsip{\rho \otimes \hat{\sigma}}{C} = \hsip{\rho}{C_\A}$ where $C_\A = \Tr_\B[(\id \otimes \hat{\sigma})C] \in \Herm(\A)$  
whose solution is the principal eigenvector of $C_\A$. 
Thus, the see-saw algorithm is simply a sequence of principal eigenvector computations. 
 
\subsection{Special-case instances.} \label{sec:special}

If $\dim(\A) = \dim(\B) = d$, then $C \in \Herm(\A : \B)$ is a $d^2 \times d^2$ matrix.  
Therefore solving $\hsip{\rho \otimes \sigma}{C}$ for either $\rho$ or $\sigma$ requires dealing with a $d^2 \times d^2$ matrix. 
Calculating $C_\A$ and $C_\B$ using a brute force method involves matrix computations on a $d^2 \times d^2$ matrix, which can grow expensive rather quickly. 
Before discussing our dimension-reduction technique, we discuss a special case that can speed up computations at this level. 

We now consider instances with a particular structure that circumvents the pre-computation issue raised above while still capturing many applications of interest. 
Suppose $C$ can be written in the following way 
\begin{equation} \label{eq:special}
    C = \sum_{m=1}^N K_m \otimes L_m,
\end{equation}
where $K_m \in \Herm(\A)$ and $L_m \in \Herm(\B)$, for all $m$. 
Observe that
\begin{equation}
    C_\B = \Tr_\A[(\hat{\rho} \otimes \id)C] = \Tr_\A \left[ (\hat{\rho} \otimes \id) \left( \sum_m K_m \otimes L_m \right) \right] = \sum_m \hsip{\hat{\rho}}{K_m} L_m 
\end{equation}
and similarly  
    $C_\A = \Tr_\B[(\id \otimes \hat{\sigma})C] = \sum_m \hsip{\hat{\sigma}}{L_m} K_m$.   
Therefore, the see-saw pre-computation involves a series of matrix computations on matrices of size $d \times d$ now, offering a quadratic speed up. 


\subsection{Numerical tests} 
  
\paragraph{Efficiency.}  
Using the approach for general $C$, we can perform see-saw up to local dimension of $30$ (which for our numerical experiments take roughly $30$ hours). 
However for instances with the structure in \eqref{eq:special}, we could solve for much larger dimensions. 
For this, we generated random $C$ matrices of the form $C = \sum_{m=1}^{10} K_m \otimes L_m$ and with this we observe that our experiments for a local dimension of $30$ takes less than $0.008$ seconds.
Moreover we are able to perform see-saw for instances with this specific structure up to a local dimension of $2500$.
At this point, our numerical experiments required about $35$ days of runtime.
(The hardware specifications are indicated in the computational platform section (Section~\ref{sec:platform}).)  

\paragraph{Accuracy.}  
    In our numerical experiments, we compared the lower bounds against some outer approximations obtained by relaxing the set $\SepD$ into a larger set that can be represented as an SDP (positive partial transpose criterion~\cite{peres1996separability, horodecki1996necessary} and symmetric extensions~\cite{doherty2002SymExt}).
    To quantify the performance, we track the difference $\beta - \gamma$ where $\beta$ is the least upper bound (e.g., from all of the outer approximations calculated) and $\gamma$ is the greatest lower bound (e.g., by performing see-saw over the various starting points). 
    If the difference is small, then we have a good approximation and the states outputted from the best see-saw algorithm is an approximately optimal solution. 
    A large difference indicates that we have found a tricky instance over which at least one of these algorithms failed to produce a good approximation. 
      
    For varying dimensions, we ran this over $100$ random instances.
    Our numerical experiments showed that this difference was at most $10^{-3}$ for values of local dimension up to $13$.
    Beyond this value of the local dimension, the median time required to compute the upper bounds over $100$ random instances of $C$ exceeded $30$ minutes and thus had to be terminated.
    Since the see-saw algorithm is seen to be effective both in terms of performance as well as scalability, we rely on see-saw for approximations for larger dimensions in the rest of this work. 

\paragraph{Data.}   
    The data collected from these numerical experiments is hosted on an interactive web application \url{https://ankith-mohan.shinyapps.io/SEP_app}. 
      
 
\section{Reducing the dimension of separability problems via a quantum co-processor}\label{sec:big_sep} 

Now that we have argued that the see-saw algorithm provides good inner approximations on small separability problems, what can we do with large separability problems? 
An even tougher problem: 
\begin{quote} 
\centering
\textit{What if the problem is so big we cannot even write it down? }
\end{quote} 
Indeed, many quantum problems suffer from the curse of dimensionality meaning that classical descriptions could involve exponentially large vectors and matrices. 
We now consider heuristics for reducing the dimension by generalizing the NISQ SDP solver idea in~\cite{bharti2022noisy} to optimize over both separable operators and separable states.   
 
\subsection{General linear optimization over separable operators} 
Consider the optimization problem  
\begin{equation}
    \mu = \text{maximize } \Big\{ \hsip{X}{C} : \Xi(X) = B, \; X \in \Sep(\A : \B) \Big\}.  
\end{equation} 
Suppose we are given \emph{two} sets of ansatz states $\{\ket{\psi_1}, \dots, \ket{\psi_L}\} \subset \A$ and $\{\ket{\phi_1}, \dots, \ket{\phi_M}\} \subset \B$. 
From these states, we define the two matrices 
\begin{equation}
    \Psi \coloneqq \sum_{i=1}^L \ketbra{\psi_i}{i} \quad \text{and} \quad \Phi \coloneqq \sum_{k=1}^M \ketbra{\phi_k}{k}.
\end{equation} 
Then we can \emph{guess}\footnote{An optimal solution may not have the following form in general, however this does yield a feasible solution. 
Intuitively, the better the ansatz, the better feasible solution one would expect to obtain.} that an optimal solution takes the form 
\begin{equation} 
X = (\Psi \otimes \Phi) Y (\Psi \otimes \Phi)^* 
\end{equation} 
for some $Y \in \Sep(\Cspace^L : \Cspace^M)$.   
Now consider the following separability problem 
\begin{equation}
    \label{eq:reduced_cone}
    \begin{aligned}
        \mu' = \text{maximize } \Big\{ \hsip{Y}{D} : \Xi'(Y) = B, \; Y \in \Sep(\Cspace^L : \Cspace^M) \Big\}, 
    \end{aligned}
\end{equation}
where 
$D = (\Psi \otimes \Phi)^* C (\Psi \otimes \Phi)$ 
and 
$\Xi'(\cdot) = \Xi((\Psi \otimes \Phi) \cdot (\Psi \otimes \Phi)^*)$.   
It can be verified that 
\begin{itemize} 
\item $\hsip{X}{C} = \hsip{Y}{D}$, 
\item $\Xi(X) = \Xi'(Y)$, and 
\item $X \in \Sep(\A : \B)$ when $Y \in \Sep(\Cspace^L : \Cspace^M)$.  
\end{itemize}  
To see the last claim, if we take a separable decomposition of $Y = \sum_i q_i \rho_i \otimes \sigma_i$, we have that 
\begin{equation} 
X 
= (\Psi \otimes \Phi) Y (\Psi \otimes \Phi)^* 
= \sum_i q_i \Psi \rho_i \Psi^* \otimes \Phi \sigma_i \Phi^* \in \Sep(\A : \B). 
\end{equation} 
Moreover, this gives an \emph{inner approximation} for the original separability problem; if $Y$ is feasible with objective function value $\hsip{Y}{D}$ in \eqref{eq:reduced_cone}, then $X$ is feasible in the original problem with objective function value $\hsip{X}{C} = \hsip{Y}{D}$. 
Thus, $\mu' \leq \mu$, yielding a bound on the optimal value of the original problem as well. 
We call the problem~(\ref{eq:reduced_cone}) the \emph{reduced} problem since, typically, its size is much smaller than the original problem. 

If we apply the see-saw algorithm to the reduced problem, this yields a feasible solution to the reduced problem. 
As we have previously discussed, the see-saw algorithm performs well on smaller separability problems. 
Moreover, this feasible solution can be turned into a feasible solution for the original problem. 
Thus, overall this process yields a feasible solution to the original problem, and this provides a lower bound on its value. 
  

\subsection{Optimizing over separable states}\label{sec:sep_states_SDP}   
We now consider the special case when the constraint $\Xi(X) = B$ is written as $\Tr(X) = 1$, i.e., we wish to optimize over separable quantum states such as in the following problem:  
\begin{equation} \label{first}
    \text{maximize } \Big\{ \hsip{\rho}{C} : \rho \in \SepD(\A:\B) \Big\}.
\end{equation} 
From the previous discussion, the reduced problem takes the form 
\begin{equation}
    \label{eq:reduced_SepD_intro}
    \text{maximize } \Big\{ \hsip{Y}{D} : \hsip{G_\A \otimes G_\B}{Y} = 1, \; Y \in \Sep(\Cspace^L : \Cspace^M) \Big\},
\end{equation}
where $G_{\A}$ is the Gram matrix of the first set of ansatz states and $G_{\B}$ is the Gram matrix of the second set of ansatz states: 
\begin{equation}
    G_\A = \sum_{i=1}^L \sum_{j=1}^L \braket{\psi_i}{\psi_j} \ketbra{i}{j} \quad \text{and} \quad G_\B = \sum_{k=1}^M \sum_{l=1}^M \braket{\phi_k}{\phi_l} \ketbra{k}{l}. 
\end{equation} 
If we suppose that the each set of ansatz states are linearly independent, then both $G_{\A}$ and $G_{\B}$ are invertible. 
Now, if we define the matrix 
\begin{equation}
    Z \coloneqq \left( G_\A^{1/2} \otimes G_\B^{1/2} \right) Y \left( G_\A^{1/2} \otimes G_\B^{1/2} \right) 
\end{equation}
we can rewrite the reduced problem as the following: 
\begin{equation}
    \label{eq:ultra_reduced_SepD_intro}
    \text{maximize } \Big\{ \hsip{Z}{\tilde{D}} : Z \in \SepD(\Cspace^L : \Cspace^M) \Big\},
\end{equation}
where 
\begin{equation} \label{Dyo}
\tilde{D} = (G_{\A}^{-1/2} \Psi^* \otimes G_{\B}^{-1/2}\Phi^*)\ C\ (\Psi G_{\A}^{-1/2} \otimes \Phi G_{\B}^{-1/2}). 
\end{equation} 
To verify this, we can see that: 
\begin{itemize} 
\item $Y$ is separable if and only if $Z$ is separable, 
\item $\hsip{Z}{\tilde{D}} = \hsip{Y}{D}$, and 
\item $\Tr(Z) = 1$ if and only if $\hsip{G_\A \otimes G_\B}{Y} = 1$. 
\end{itemize} 

Let us summarize what we have accomplished so far. 
We have taken the optimization over separable states given in \eqref{first} defined by the matrix $C \in \SepD(\A : \B)$ and replaced it with a reduced version which is itself an optimization over separable states given in \eqref{eq:ultra_reduced_SepD_intro} defined by the matrix $\tilde{D} \in \SepD(\Cspace^L : \Cspace^M)$.  
This can be a considerably smaller optimization over separable states and, again, given a feasible solution to the reduced problem we can find a feasible solution to the original problem. 


\subsection{Calculating the data of the reduced problems with a quantum co-processor}    
 
\begin{figure}[ht]
    \centering
    \begin{quantikz}
        \lstick{$\ket{0}$} & \gate{H} & \ctrl{1} & \gate{S^b} & \gate{H} & \meter{} & \cw \\
        \lstick{$\ket{\psi}$} & \qw & \gate{U} & \qw & \qw & \qw & \qw \\
    \end{quantikz}    
    \caption{
    The circuit for the Hadamard test which approximates the inner product between two states $\ket{\psi}$ and $\ket{\phi}$ given many samples. 
    Here, the unitary $U$ maps the state $\ket{\psi}$ to the state $\ket{\phi}$.
    When we set $b = 0$, the circuit approximates  $\Re(\braket{\psi}{\phi})$, and when we set $b = 1$,  it approximates $\Im(\braket{\psi}{\phi})$. 
    }
    \label{fig:Hadamard_test}
\end{figure}
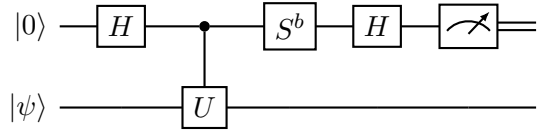

One immediate concern arises when one wishes to compute $\tilde{D}$. 
This involves computing matrix operations on possibly huge vectors, which may not be possible on a classical computer. 
This is where the quantum co-processor comes into the picture. 
To this end, we make the assumption that we have physical access to the ansatz states $\ket{\psi_i}$ and $\ket{\phi_j}$. 
From an algorithmic standpoint, we assume that they are efficiently preparable and that we have access to: 
\begin{itemize} 
\item circuits $U_i$ that create $\ket{\psi_i}$ and their inverses, and 
\item circuits $V_j$ that create $\ket{\phi_j}$ and their inverses. 
\end{itemize} 
By combining them, we thus have a circuit $U$ which maps $\ket{\psi_i}$ to $\ket{\psi_j}$ for any choice of $i$ and $j$, and similarly for the $\ket{\phi_i}$ states.  
This way we can compute the Gram matrix $G_{\A}$ by performing the \emph{Hadamard test} (Figure~\ref{fig:Hadamard_test}) to learn $\mathrm{Re}(\braket{\psi_i}{\psi_j})$ and $\mathrm{Im}(\braket{\psi_i}{\psi_j})$ for each pair of $i$ and $j$. 
We can similarly compute the Gram matrix $G_{\B}$. 
 
In the event that all of the inner products are non-negative then we could alternatively use the \emph{SWAP test} to compute each of the inner products in the Gram matrix. 
In this setting, we do not require any of the preparation circuits, just access to copies of the states. 

Lastly, as long as the ansatz states are linearly independent and there are not too many of them, we can relatively easily compute the inverses of their Gram matrices. 

With this said, all that remains is to compute $\tilde{D}$, which we now discuss. 
Recall that we have $\tilde{D} = \left( G_{\A}^{-1/2} \otimes G_{\B}^{-1/2} \right)\ D\ \left( G_{\A}^{-1/2} \otimes G_{\B}^{-1/2} \right)$, and we can write  
\begin{equation}
    D := \sum_{i=1}^L \sum_{j=1}^L \sum_{k=1}^M \sum_{l=1}^M \bra{\psi_i}\bra{\phi_k} C \ket{\psi_j} \ket{\phi_l} \ketbra{i}{j} \otimes \ketbra{k}{l}. 
\end{equation} 
Thus, if we can compute $D$ via a quantum co-processor, then we are done. 

If we were to write $C = \sum_{m} P_m$ as a linear combination of Pauli matrices, and we also have access to Pauli gates, then we can compute each entry of $D$ in a manner similar to that described above. 
That is, we can apply each $P_m$ to each the states $\ket{\psi_j}\ket{\phi_l}$ and calculate the required inner products via the Hadamard or SWAP test, say.

As a demonstration, a particularly nice case is when each of the ansatz states 
$\ket{\psi_i}$ can be prepared as $P^A_i \ket{\psi}$ where $\ket{\psi}$ is a fixed reference state, and $\ket{\phi_k}$ can be prepared as $P^B_k \ket{\phi}$ where $\ket{\phi}$ is a fixed reference state, 
where each $P \in \{\id, X, Y, Z\}^{\otimes m}$ is a Pauli string.  
Then, each entry of the matrix $D$ can be expressed as 
\begin{equation} 
D_{i,j,k,l} = \bra{\psi} \bra{\phi} \left( \sum_{m} (P^A_i \otimes P^B_k) P_m (P^A_j \otimes P^B_l) \right) \ket{\psi} \ket{\phi} = \bra{\psi} \bra{\phi} P \ket{\psi} \ket{\phi}, 
\end{equation} 
where $P$ is simply a sum of Pauli strings (with $\{\pm 1, \pm i\}$ coefficients). 
So calculating $D_{i,j,k,l}$ is equivalent to computing the expectation value of a Pauli string with respect to the state $\ket{\psi}\ket{\phi}$, which can be done efficiently as long as $\ket{\psi}\ket{\phi}$ are efficiently preparable as discussed above and we have access to Pauli gates.  


\subsection{A special case which speeds up the see-saw algorithm when applied to the reduced problem}   
We note that when $C$ can be expressed as $C = \sum_m K_m \otimes L_m$, we can rewrite the reduced problem~(\ref{eq:ultra_reduced_SepD_intro}) as
\begin{equation}
    \label{eq:ultra_reduced_SepD_KL_intro}
    \text{maximize} \left\{ \left\langle Z, \sum_{m=1}^N \tilde{K}_m \otimes \tilde{L}_m \right\rangle : Z \in \SepD(\Cspace^L : \Cspace^M) \right\},
\end{equation}
where 
$\tilde{K}_m = G_{\A}^{-1/2} \Psi^*\ K_m\ \Psi G_{\A}^{-1/2}$
and 
$\tilde{L}_m = G_{\B}^{-1/2} \Phi^*\ L_m\ \Phi G_{\B}^{-1/2}$.  
Depending on the nature of each $K_m$ and $L_m$ and our choices of ansatz states, we could have $\tilde{K}_m$ and $\tilde{L}_m$ easily computable on a quantum co-processor. 
Moreover, the structure is the same as discussed in Section~\ref{sec:special}, meaning that we can dramatically speed up the see-saw algorithm when applied to this problem. 


\section{Applications to finding separable ground states of Hamiltonians}\label{sec:Hammy_intro} 

In this section, we use see-saw and our dimension-reduction idea to solve for the separable state of the greatest\footnote{Note that since we are maximizing throughout the discussions in this work, we think of the largest eigenvalue as the ground energy. 
This convention is without loss of generality since one can replace the Hamiltonian $H$ with $-H$ if one prefers to minimize.} energy of a certain Hamiltonian. 
The discussion in this section is broken into several parts. 
The first part introduces a \emph{measure of how much entanglement} is in the ground space of a Hamiltonian. 
The second part discusses how one can approach choosing the ansatz states to run the dimension-reducing heuristic. 
We also graph how well the heuristics perform with respect to the number of ansatz states chosen. 
The third part compares how well our heuristic performs compared to the (a) full see-saw algorithm, (b) full ground energy calculation, and (c) the heuristic ground energy calculation from~\cite{bharti2022noisy}. 
We run all of our tests on the one-dimensional Ising Hamiltonian, noting that our methods apply to many other popular Hamiltonians as well. 


\subsection{A ground space entanglement measure} 

For a Hamiltonian $H$, let us denote the largest and smallest eigenvalues by $\lambda_{\mathrm{max}}$ and $\lambda_{\mathrm{min}}$, respectively.  

\begin{definition}[Separable ground energy] 
The separable ground energy of a Hamiltonian $H$ is the optimal value of the following optimization problem 
\begin{equation} \label{ThanosSnapHammy}
    \alpha = \text{maximize } \Big\{ \hsip{\tau}{H} : \tau \in \SepD(\A:\B) \Big\}. 
\end{equation} 
 
\end{definition} 

Since we cannot always efficiently compute the exact value of $\alpha$, due to the intrinsic difficulty of optimizing over separable states, we typically rely on approximations of $\alpha$, denoted $\hat{\alpha}$.  
 
Using these values, we define a measure of entanglement as the (normalized) difference between the ground energy (i.e., $\lambda_{\max})$ and $\alpha$, and an approximate version using $\hat{\alpha}$.  
Roughly speaking, we define a measure of how close a separable state is to the ground space. 
 
\begin{definition}[Ground space entanglement measure, and an approximation]
\label{def:groundspace_ent}
For a non-zero Hamiltonian $H$, we define a measure of ground space entanglement as 
\begin{equation}
    \delta = \frac{\lambda_{\mathrm{max}} - \alpha}{\lambda_{\mathrm{max}} - \lambda_{\mathrm{min}}} \in [0,1].
\end{equation}  
We note that $\delta = 0$ if and only if the ground space contains a separable state (even though it can also contain entangled states as well).   
When any of the above quantities are not known, we define an approximate ground space entanglement measure as 
\begin{equation}
    \hat{\delta} = \frac{\widehat{\lambda}_{\mathrm{max}} - \hat{\alpha}}{\widehat{\lambda}_{\mathrm{max}} - \widehat{\lambda}_{\mathrm{min}}} 
\end{equation}  
where the hats represent the best approximations to the actual values that we can obtain.\footnote{Sometimes we can calculate $\lambda_{\max}$ and $\lambda_{\min}$ easily, and for large instances one can approximate them using the heuristic eigenvalue solver in~\cite{bharti2022noisy}.} 
We note that we cannot place bounds on $\hat{\delta}$ due to the inability to place general bounds on $\hat{\alpha}$ relative to $\alpha$ most of the time. 
\end{definition} 

At first glance, it might be tempting to define the ground space entanglement as quantifying the amount of entanglement in a ground state. 
While this could have its applications, we have chosen a different definition, which we now motivate with an illustrative example. 
Consider the following two-qubit Hamiltonian
\begin{equation}\label{eq:special_Hammy}
    H = (1 - \varepsilon) \ketbra{00}{00} + \ketbra{\psi^-}{\psi^-},
\end{equation}
where $\ket{\psi^-} = \frac{1}{\sqrt{2}}(\ket{01} - \ket{10})$ and $\varepsilon > 0$ is a small constant. 
For any $\varepsilon$, the (unique) ground state of $H$ is $\ket{\psi^-}$ which is a maximally entangled state (to which most entanglement measures would assign a nontrivial value). 
However, we have $\lambda_{\max} = 1$, $\lambda_{\min} = 0$, $\alpha \geq 1-\varepsilon$, and thus $\delta \leq \varepsilon$. 
Therefore, our measure is small, indicating that there is a separable state that has energy \emph{close} to that in the ground space. 


\subsubsection{The Ising Hamiltonian} 

We consider the one-dimensional Ising model of $N$ qubits with transverse field $h$, longitudinal field $g$, and coupling term $J$, defined as
\begin{equation}
    H_{\mathrm{Ising}} = H_{\mathrm{z}} + H_{\mathrm{x}} = - \sum_{n=1}^N [J \sigma_n^z \sigma_{n+1}^z + g \sigma_n^z + h \sigma_n ^x]
\end{equation}
where $H_z = - \sum_n [J \sigma_n^z \sigma_{n+1}^z + g \sigma_n^z]$ and $H_x = -\sum_n h \sigma_n^x$, and $\sigma^z$ and $\sigma^x$ denote the Pauli $Z$ and $X$ matrices, respectively.
The shorthand notation $\sigma_n^z$ indicates that in the sequence of $N$ qubits, the Pauli $Z$ matrix is affecting the $n$th qubit while the remaining qubits are left alone.
For example, $\sigma_3^z := \id_1 \otimes \id_2 \otimes \sigma^z \otimes \id_4 \otimes \cdots \otimes \id_N$. 

\begin{figure}[htpb]
    \centering
    \subfloat[\centering $12$ qubits total]{{\includegraphics[scale=0.435]{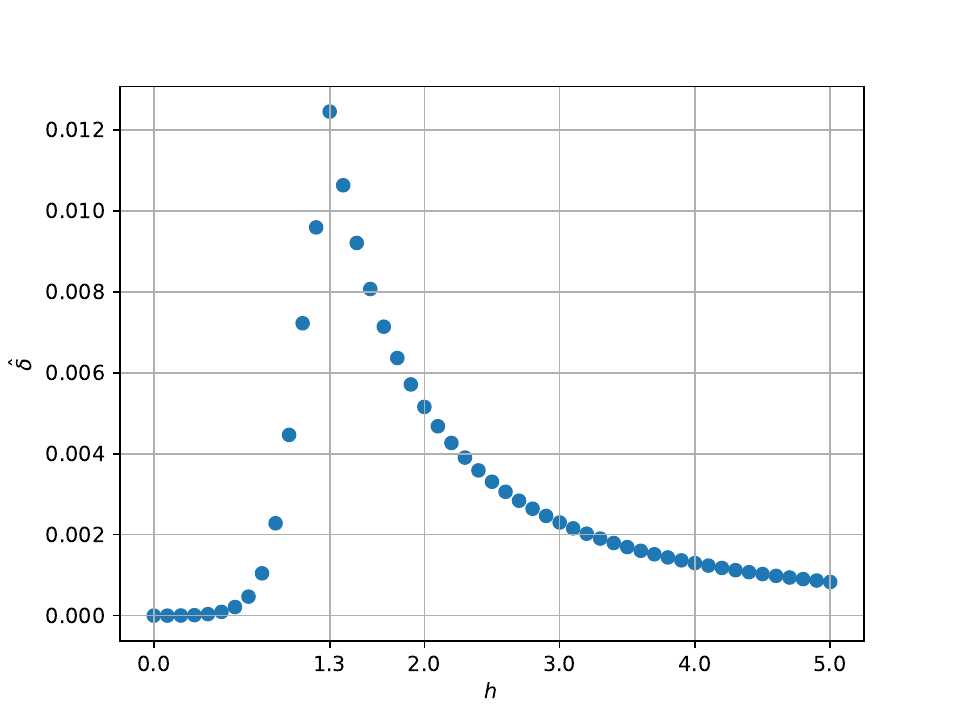} }}%
    \subfloat[\centering $14$ qubits total]{{\includegraphics[scale=0.435]{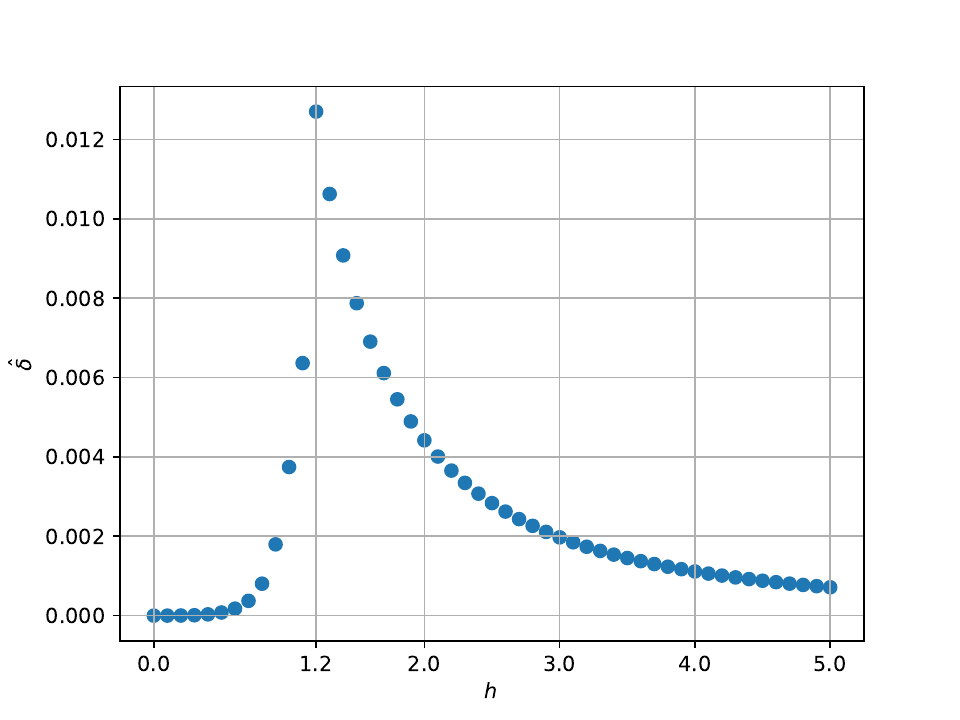} }}%
    \caption{
        A depiction of the approximate ground space entanglement measure as a function of the transverse field $h$, for $H_{\mathrm{Ising}}$ with $12$ qubits (left) and $14$ qubits (right) setting $J = 1$ and $g = 0$. 
        Here we set $\A$ to be the first half of the qubits and $\B$ to be the second half of the qubits. 
        Note that we are able to compute $\lambda_{\max}$ and $\lambda_{\min}$ for these graphs. 
    }
    \label{fig:delta_vs_h_intro}
\end{figure}

For this Hamiltonian with parameters $J$, $g$, and $h$, we first want to know at what values of these parameters will the ground space entanglement measure (i.e., $\hat{\delta}$) peak.
To answer this, we set $J = 1$, $g = 0$, and vary the value of $h$ in the range $[0, 5]$. 
Figure~\ref{fig:delta_vs_h_intro} illustrates our entanglement measure for $12$ and $14$ qubits. 

We observe that for the $12$-qubit case, $\hat{\delta}$ peaks around $h = 1.3$ and for the $14$-qubit case the peak is around $h = 1.2$. 
These graphs suggest that there is a non-trivial amount of entanglement in these ground spaces. 
Note that while this model has a phase transition for the critical point $h=1$, the entanglement is actually maximal at a value that is close, but not exactly at the critical point~\cite{osterloh2002scaling}, which matches our results. 
Going forward, we set $h$ to be in this range and test other figures of interest.


\subsection{Evaluating the performance of our heuristic with varying numbers of ansatz states} 

We now study the performance of our heuristic for reducing the dimensionality of large separability problems discussed in Section~\ref{sec:big_sep} for the Hamiltonian $H_{\mathrm{Ising}}$.
To formulate our problem as an instance of~(\ref{eq:ultra_reduced_SepD_KL_intro}), we require two things: (a) the ansatz states $\{\ket{\psi_1}, \dots, \ket{\psi_L}\} \subset \A$ and $\{\ket{\phi_1}, \dots, \ket{\phi_M}\} \subset \B$, and (b) the matrices $K_m$ and $L_m$ for all values of $m$.

\subsubsection{Choosing the ansatz states}

We use the \emph{NISQ-friendly version of the Krylov subspace} approach proposed by~\cite{bharti2022noisy} to generate the ansatz space $\Sset$.
The original Krylov subspace approach approximates the ground state of the Hamiltonian $H$ up to order $K$, using a reference state $\ket{\eta}$ as 
\begin{equation}
    \ket{\xi(\chi)}^{(K)} = \chi_0 \ket{\eta} + \chi_1 H \ket{\eta} + \dots + \chi_K H^K \ket{\eta} 
\end{equation} 
where $\chi$ is a $(K+1)$-dimensional vector.
If a Hamiltonian can be expressed as a weighted sum of Pauli strings, i.e., $H = \sum_m c_m P_m$, then $H^k$ can be expressed as 
\begin{equation} 
H^k = \left( \sum_m c_m P_m \right)^k = \sum_{m_1, \dots, m_k} c_{m_1, \dots, m_k}\ P_{m_1} P_{m_2} \cdots P_{m_k}, 
\end{equation} 
which can be further rewritten as the weighted sum of products of Pauli strings.
Each product of Pauli strings applied to the reference state, $P_{m_1} \cdots P_{m_k}\ket{\eta}$, is a state that we consider to be added to our set of ansatz states. 
This process repeated up to power $K$ results in an ansatz space that contains the original Krylov subspace. 
The first order of the ansatz states for $H_{\mathrm{Ising}}$ is as follows: 
\begin{equation}
    \label{eq:OG_Krylov}
    \{\sigma_1^z \ket{\eta}, \sigma_2^z \ket{\eta}, \dots, \sigma_N^z \ket{\eta}, 
                    \sigma_1^z \sigma_2^z \ket{\eta}, \sigma_2^z \sigma_3^z \ket{\eta}, \dots, \sigma_N^z \sigma_1^z \ket{\eta},
                    \sigma_1^x \ket{\eta}, \sigma_2^x \ket{\eta}, \dots, \sigma_N^x \ket{\eta}\}.
\end{equation}
  
In our case, we generate the ansatz states $\{\ket{\psi_1}, \dots, \ket{\psi_L}\} \subset \A$ and $\{\ket{\phi_1}, \dots, \ket{\phi_M}\} \subset \B$, in the context of $H_{\mathrm{Ising}}$, by picking the reference states $\ket{\psi} \in \A$ and $\ket{\phi} \in \B$, and using the NISQ-friendly version of the Krylov subspace approach. 
We construct $\Sset_\A$ by first generating the states in~\eqref{eq:OG_Krylov} with Pauli strings that act trivially on $\B$ with a reference state $\ket{\psi} \in \A$ (so that the states are understood to be in $\A$) and choosing $L$ linearly independent states from this set. 
We construct $\Sset_\B$ analogously, by considering Pauli strings that act trivially on $\A$, then choosing $M$ linearly independent states.


\subsubsection{Representing the Ising Hamiltonian in the form $\sum_m K_m \otimes L_m$}  
To obtain the matrices $K_m$ and $L_m$ to represent the Ising Hamiltonian, suppose that of the $N$ qubits, $N_\A$ of them correspond to the first set of ansatz states, while the remaining $N_\B$ correspond to the second set of ansatz states, i.e., $N_\A + N_\B = N$.
Then 
\begin{equation}
    \begin{aligned}
        -\sum_{n=1}^N J &\sigma_n^z \sigma_{n+1}^z = \\
        & \sqrt{J} H_\A^{zz} \otimes \sqrt{J} \id_\B +
        \sqrt{J} \id_\A \otimes \sqrt{J} H_\B^{zz} + 
        \sqrt{J} \sigma_{N_\A}^z \otimes \sqrt{J} \sigma_{{N_\A}+1}^z + 
        \sqrt{J} \sigma_1^z \otimes \sqrt{J} \sigma_N^z,
    \end{aligned}
\end{equation}
where
        $H_\A^{zz} = -\sum_{l=1}^{{N_\A}-1} \sigma_l^z \sigma_{l+1}^z$ represents the pairwise terms contained only on the first $N_\A$ qubits, 
        $H_\B^{zz} = -\sum_{m={N_\A}+1}^{N-1} \sigma_m^z \sigma_{m+1}^z$ represents the pairwise terms contained only on the remaining $N_\B$ qubits, and the last two terms above represent the two terms which act non-trivially on qubits $N_\A, N_\A +1$ and on qubits $1, N$.   
Also, we have 
\begin{equation}
    \begin{aligned}
        -\sum_{n=1}^N g \sigma_n^z = \sqrt{g} H_\A^z \otimes \sqrt{g} \id_\B + 
        \sqrt{g} \id_\A \otimes \sqrt{g} H_\B^z,
    \end{aligned}
\end{equation}
where
        $H_\A^z = -\sum_{l=1}^{N_\A} \sigma_l^z$ and 
        $H_\B^z = -\sum_{m=N_\A+1}^N \sigma_m^z$ represent the single-qubit $\sigma^z$ terms. 
Lastly,
\begin{equation}
    \begin{aligned}
        -\sum_{n=1}^N h \sigma_n^x = \sqrt{h} H_\A^x \otimes \sqrt{h} \id_\B + 
    \sqrt{h} \id_\A \otimes \sqrt{h} H_\B^x,    
    \end{aligned}
\end{equation}
where
        $H_\A^x = -\sum_{l=1}^{N_\A} \sigma_l^x$ and 
        $H_\B^x = -\sum_{m={N_\A}+1}^N \sigma_m^x$ represent the single-qubit $\sigma^x$ terms. 
        
\subsubsection{Varying the number of ansatz states} 
 
We now address the following question: How does the number of ansatz states $L$ and $M$ affect the performance of our heuristic? 
To answer this, we consider the $12$-qubit Ising Hamiltonian (with $J = 1, g = 0$ and $h = 1.3$) and use the see-saw algorithm for special-case instances discussed in Section~\ref{sec:faster_see-saw} to compute the lower bound $\hat{\alpha}$. 
We now compare $\hat{\alpha}$, which is our approximation to the actual separable ground energy, to the value attained if we were to use our dimension-reducing ansatz. 
We note that while we do not need to use dimension reduction here, it serves as a test bed to see how well we could expect it to perform on larger examples where we might not know the answer. 

\begin{figure}[htpb]
    \centering
    \includegraphics[scale=0.6]{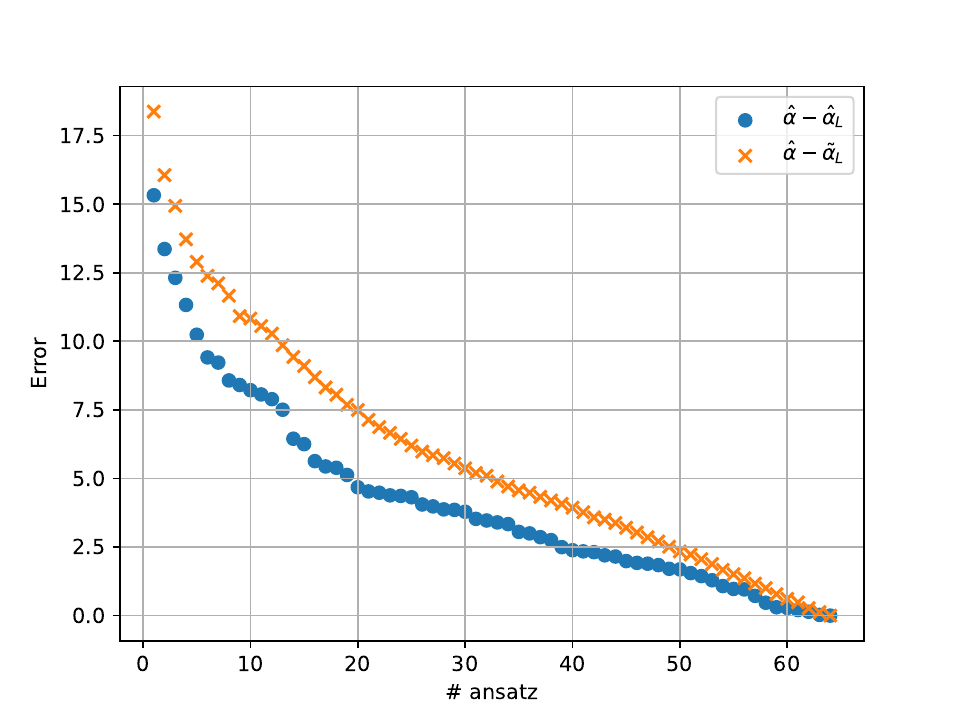}
    \caption{
        The error generated by the use of our heuristic as we vary the number of ansatz states for the Ising Hamiltonian on 12 qubits (with $J = 1, g = 0, h = 1.3$). 
        Here $\hat{\alpha}_L$ and $\tilde{\alpha}_L$ denote the maximum and the mean of the $\hat{\alpha}_{L,i}$ values, respectively, computed using $10$ randomly chosen reference states. 
    }
    \label{fig:vNa_6qe_intro}
\end{figure}

Suppose that these $12$ qubits are split evenly between Alice and Bob and we have up to $2^6 = 64$ ansatz states each.   
To begin, we randomly choose a reference state $\ket{\eta}$ then generate $L \in \{ 1, \ldots, 64 \}$ ansatz states for each of the varying values of $L$. 
More precisely, we pick the first $L$ and numerically verify that they are linearly independent so that the inverses of their Gram matrices are well-defined. 
Since the reference state is chosen randomly, we repeat this $10$ times for each $L$. 
We compute the lower bound, denoted $\hat{\alpha}_{L,i}$, for the $i$th choice of reference state for the choice of $L$. 
We denote the mean of these $10$ values as $\tilde{\alpha}_L$ and its maximum by $\hat{\alpha}_L$. 
Figure~\ref{fig:vNa_6qe_intro} describes the difference between $\hat{\alpha}$ and these values, for each value of $L$. 

This gives an illustration of how many ansatz states one could choose to use. 
We observe that the error decreases to near $0$ as the number of ansatz states increases. 
However, the computation gets more demanding as the number of ansatz states increases, so there is a trade-off. 
From this numerical experiment, it seems that the errors decrease at a slower rate starting around $L = 20$. 


\subsection{Calculating ground energies and separable ground energies on large Ising Hamiltonians}

In this subsection, we consider the $2N$-qubit Ising Hamiltonian with $J = 1$, $g=0$, and $h = 1.4$ where $\A$ is the first $N$ qubits and $\B$ is the remaining $N$ qubits. 
Our computations here are split into two camps: 
\begin{itemize} 
\item 
First, we compute the separable ground energy approximation $\hat{\alpha}$, when possible, and also our heuristic approximation $\hat{\alpha}_L$ as discussed in the previous subsection, using the maximum number of ansatz states $L$ that would fit into memory, computed over $10$ randomly chosen reference states. 
\item Second, we compute the ground energy $\lambda_{\mathrm{max}}$ using an eigenvalue solver, when possible, and $\hat{\lambda}_{\mathrm{max}}$ using the approximate eigenvalue solver in~\cite{bharti2022noisy}.
\end{itemize} 

\begin{figure}[htpb]
    \centering
    \includegraphics[scale=0.6]{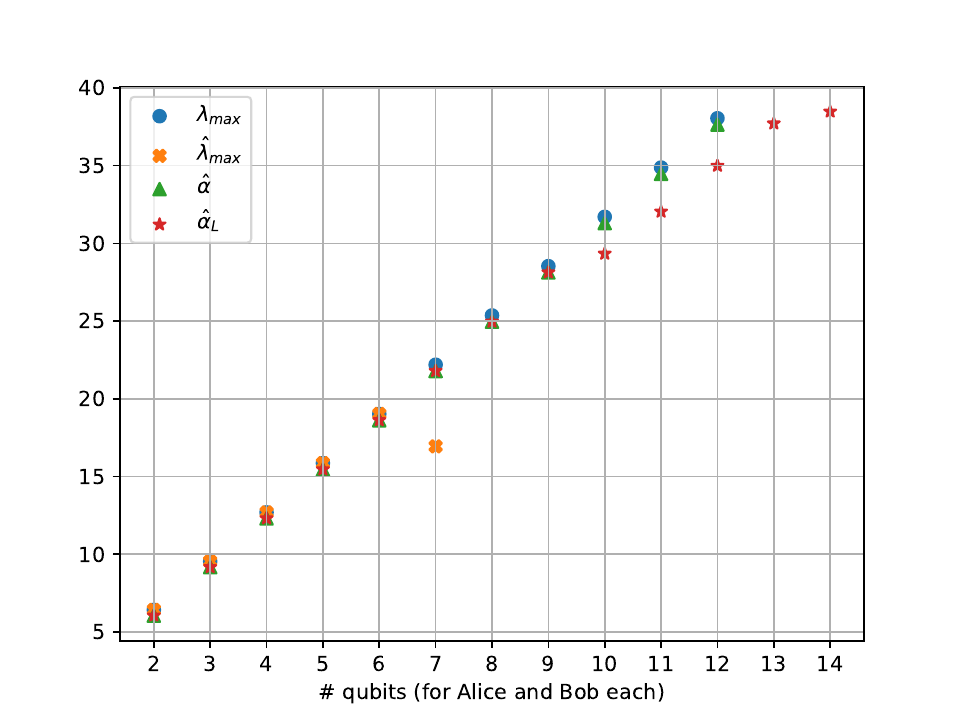}
    \caption{
        Comparing the calculations (when possible) and approximations to the ground energy and separable ground energy for the Ising Hamiltonian ($J = 1, g = 0, h = 1.4$) for up to $28$ qubits. 
    }
    \label{fig:NSeS_vs_NSS_intro}
\end{figure}

The values are shown in Figure~\ref{fig:NSeS_vs_NSS_intro}. 

We note there are are several points of interest in the above figure. 
First, at $N = 7$ we observe that there is a gap between $\hat{\lambda}_{\max}$ and $\lambda_{\max}$ noting that this is also a heuristic-based approach to computing eigenvalues. 
After this value of $N$, we stopped computing this heuristic as it became too expensive.  

For $N=9$, we can store an entire basis of ansatz states while for $N \geq 10$, we do not have sufficient memory to store enough ansatz states to choose a basis. 
Since we do not have a basis, we are solving a smaller dimensional optimization problem and therefore we see a small dip in performance like in Figure~\ref{fig:vNa_6qe_intro}.

When $N \geq 13$, the computation of our inner approximation $\hat{\alpha}$ (without ansatz) exceeds $24$ hours and therefore had to be terminated. 
However, the approximation $\hat{\alpha}_L$ is still able to be calculated, and we do so for the largest possible $L$ for each number of qubits. 
Also, $\lambda_{\max}$ can be computed up to $N = 12$, noting that this is a $2N$-qubit Hamiltonian and thus  
this computation involves computing the eigenvalue of a $2^{24} \times 2^{24}$ matrix. 
Despite the sparsity of $H_{\mathrm{Ising}}$, for values of $N \geq 13$, we are unable to compute the value of $\lambda_{\max}$ because the size of the Hamiltonian exceeds the available memory ($32$ GB, as described in the computational platform section (in Section~\ref{sec:platform})).  

At $N = 15$, the size of each of the $K_m$ and $L_m$ matrices exceeds the available memory. 

\section{Conclusions} 

In this work, we studied SDP-based algorithms for solving small separability problems as well as a heuristic for reducing the dimension of large separability problems. 
We numerically tested the performance of these algorithms from which we concluded that the see-saw algorithm performs well on small separability problems and our heuristic-based algorithm works increasingly well as the number of ansatz states increases. 
By exploiting the structure of the one-dimensional Ising Hamiltonian, we  were able to apply our heuristics on large instances to approximate the separable ground energy for up to $28$ qubits. 

Future work can extend our methods to calculate the dynamical evolution under Hamiltonians~\cite{bharti2021quantum,haug2022generalized,lau2022nisq} as well as include the effect of symmetries to reduce problem sizes~\cite{bharti2022noisy}. 

 
\subsection*{Computational platform}\label{sec:platform}

All computations were performed on a $32$ GB $10$th Generation Intel Core i$9$-$10885$H CPU ($16$ MB cache, $2.40$ GHz, $8$ cores).


\subsection*{Acknowledgements} 

AM thanks Sophia Economou for helpful discussions. 
This work was supported by Fujitsu Research of America. KB acknowledges the support from Q.InC Strategic Research and Translational Thrust.

\bibliographystyle{alpha}
\bibliography{SEP}

\end{document}